\documentclass[aps,pra,showpacs,twoside,twocolumn,10pt,superscriptaddress,nofootinbib]{revtex4-2}
\usepackage[colorlinks=true, citecolor=red, urlcolor=blue]{hyperref}
\usepackage{mathtools,braket,soul,enumitem,color,bbold,multirow,graphicx,array}
\usepackage[normalem]{ulem}
\usepackage[table,xcdraw]{xcolor}

\usepackage[ruled]{algorithm2e}

\newcolumntype{C}[1]{>{\centering\arraybackslash}m{#1}} 

\RequirePackage{dsfont}
\newcommand{\id}{\mathds{1}}

\begin{document}

\title{Rescaling decoder for 2D topological quantum color codes on 4.8.8 lattices}
	
\author{Pedro Parrado-Rodr\'iguez}
\affiliation{%
Department of Physics, College of Science, Swansea University, Singleton Park, Swansea SA2 8PP, United Kingdom}%

	\author{Manuel Rispler}
	\affiliation{Institute for Quantum Information, RWTH Aachen University, D-52056 Aachen, Germany}
\affiliation{Peter Gr\"unberg Institute, Theoretical Nanoelectronics, Forschungszentrum J\"ulich, D-52425 J\"ulich, Germany}

	\author{Markus M{\"u}ller}
\affiliation{Institute for Quantum Information, RWTH Aachen University, D-52056 Aachen, Germany}
\affiliation{Peter Gr\"unberg Institute, Theoretical Nanoelectronics, Forschungszentrum J\"ulich, D-52425 J\"ulich, Germany}

%	\vspace{-3.5cm}

	\begin{abstract}	
		Fault-tolerant quantum computation relies on scaling up quantum error correcting codes in order to suppress the error rate on the encoded quantum states. Topological codes, such as the surface code or color codes are leading candidates for practical scalable quantum error correction and require efficient and scalable decoders. In this work, we propose and study the efficiency of a decoder for 2D topological color codes on the 4.8.8 lattice, by building on the work of~\cite{Sarvepalli_2012} for color codes on hexagonal lattices. The decoder is based on a rescaling approach, in which syndrome information on a part of the qubit lattice is processed locally, and then the lattice is rescaled iteratively to smaller sizes. We find a threshold of 6.0\% for code capacity noise.
	\end{abstract}
	
	\maketitle

\section{Introduction}\label{sec.introduction}

Quantum error correction (QEC) schemes aim at detecting and correcting errors during storage and processing of quantum information to enable long and reliable quantum computation on scalable quantum processors~\cite{nielsen-book, Terhal2015}. QEC codes encode logical information in non-local degrees of freedom such that the effects of errors can be detected through parity check measurements and reversed before they accumulate. The threshold theorem ensures that the logical error rate can be arbitrarily suppressed by increasing the size of the code, provided that fault-tolerant quantum circuit constructions are used and the physical error rates fall below a given critical threshold~\cite{Aharonov2008,Shor1996,Preskill1998}. The value of the threshold depends on the QEC code, the noise model and in particular also on the decoder. The latter amounts to our ability to correctly interpret the syndrome, which is given by the collection of error information gathered through measuring the parity checks, in order to apply a correction with a high success probability. Topological QEC codes such as the toric or surface code~\cite{Kitaev1997, Kitaev2003, Dennis2002} and color codes~\cite{colorCodes2006bombin,bombin2007colorcodes3d} encode the logical information into topological properties of the system, while all parity checks are low-weight measurements involving geometrically local qubits on the lattice. They stand out as the QEC codes with some of the highest known thresholds~\cite{Terhal2015,raussendorf-prl-98-190504}, when e.g.~compared to concatenated codes,  
 which makes them attractive candidates for experimental realizations~\cite{Nigg2014,Ryan-Anderson2021,Satzinger2021,wallraff2020surface,Postler2021,dicarlo2017surface,marques2021logicalqubit,takita2016surface}. In particular the seven qubit color code as the smallest fully-functional representative of the family of 2D color codes has been targeted in a series of experimental QEC advances in ion trap quantum devices: from code state preparation implemented in~\cite{Nigg2014}, fault-tolerant stabilizer measurement~\cite{hilder2021flagreadout} and repeated QEC cycles ~\cite{Ryan-Anderson2021} as well as fault-tolerant magic state preparation and injection~\cite{Postler2021} have been demonstrated recently. Complementary, small surface codes are prominently being pursued in superconducting qubit experiments: there, single weight-four parity measurement~\cite{takita2016surface} and error detecting surface codes~\cite{marques2021logicalqubit,wallraff2020surface}, code state preparations of larger surface code states~\cite{Satzinger2021} and the very recent leap towards repetitive QEC cycles in a surface-17 architecture~\cite{Krinner2021} have been shown.

To execute operations on the encoded logical qubits, one has to devise logical gates, which is the subject matter of the theory of fault tolerance~\cite{Shor1996,Preskill1998}. Here, the challenge is to make sure that the synthesized gates act in such a way that they cannot inadvertently spread errors beyond the scope of what the QEC code is able to correct. This spreading is most easily avoided by acting on the data qubits within the code block separately, i.e.~using no entangling operations at all, which is known as a transversal implementation of a logical gate. However, the possibility of implementing logical gates transversally is limited: on the lower end it depends on the QEC code and on the upper end it is known to be impossible to implement a universal gate set fault-tolerantly by a no-go theorem by Eastin and Knill~\cite{Eastin2009}. In this regard, the color code on the 4.8.8 lattice is particularly interesting, because it allows for the transversal implementation of the entire Clifford group, which distinguishes it from surface codes or color codes on hexagonal lattices. It is therefore optimal in the sense that adding any other (non-Clifford) gate would render the gate set universal and hence violate the no-go theorem. The remaining non-Clifford gate is typically synthesized by other means such as magic state distillation and injection~\cite{kitaev2005magic,chamberland2020ftmagic,krishna2019overhead,litinski2019cost}, for which color codes are also particularly relevant~\cite{landahl2011faulttolerant,colorCodes2006bombin}, as underlined by recent fault-tolerant implementations of Clifford gates~\cite{Ryan-Anderson2021} and a non-Clifford T-gate~\cite{Postler2021}.
 
To operate the color code, in particular codes of larger distance, it is vital to have an efficient decoding algorithm. This decoder should on the one hand perform as well as possible in terms of proposing a (near-)optimal recovery operation. This performance is reflected in the threshold value, and it can at least for simple noise models be benchmarked against known upper bounds on decoding performance obtained through mapping the quantum error correction problem onto a classical statistical-mechanical model~\cite{Dennis2002, Katzgraber_2009,katzgraber2011tricolored}. On the other hand, this accuracy of decoding has to be balanced with the time it takes to run the decoding algorithm: while for a quantum memory, it is potentially fine to keep a backlog of measured error syndromes and figure out the correction in classical postprocessing later, this is not the case anymore once we start to perform logical quantum computations: here the intermediate state of the computation will depend on the decoder outcome, such that the quantum computation may have to wait for the decoding algorithm to finish, time during which of course new errors accumulate. The development of decoders for color codes has been and still is an active field of research~\cite{Sarvepalli_2012,wang2009graphical,stephens2014efficient,Maskara2019networks,Delfosse_2014projection,delfosse2017almostlinear,Delfosse_2020peeling,Kubica2019cellular,kubica2019restriction,Baireuther_2019,Chamberland_2018,Davaasuren_2020}. For the case of data qubit noise, also known as code capacity noise, the decoder upper bound threshold is known to be 10.9\%~\cite{Katzgraber_2009}. The decoder with the best performance known in terms of threshold is the restriction decoder, which achieves 10.2\%, when using minimum weight perfect matching (MWPM) as a subroutine~\cite{kubica2019restriction}. The runtime complexity of MWPM as a function of the number of qubits $N$ is $\mathcal{O}(N^4)$ for a straightforward implementation of the original Blossom algorithm~\cite{Edmonds1965,blossomKolmogorov}, which can be optimized to $\mathcal{O}(N^{2.5})$ using recent advancements~\cite{Gabow2017,duan2018scalingWM}. Instead of MWPM, one could also use the union-find decoder~\cite{delfosse2017almostlinear}, which has a runtime complexity of $\mathcal{O}(N \alpha(N))$, where $\alpha(x)$ is the inverse Ackermann function and extremely slowly growing, which equips this decoder with almost linear time complexity. In this work, we explore a different decoding paradigm based on iterative rescaling and partial decoding known as renormalization-group (RG) decoding. This paradigm was introduced in~\cite{ducloscianci2010renormalization} for surface codes and extended to the color code on a hexagonal (6.6.6) lattice in~\cite{Sarvepalli_2012}. Owing to the rescaling feature, these algorithms have a runtime complexity of $\mathcal{O}(N \log(N))$, as the rescaling of the lattice can be done in linear time $\mathcal{O}(N)$, since it is based on local operations, and the number of iterations of the rescaling process grows as $\mathcal{O}(\log(N))$. The local nature of the rescaling algorithm means it can be parallelized, achieving an overall scaling of $\mathcal{O}(\log(N))$, which holds the potential for drastic improvements in decoding runtime.

\begin{figure}[ht!]
\begin{center}
\includegraphics[width=0.99\columnwidth]{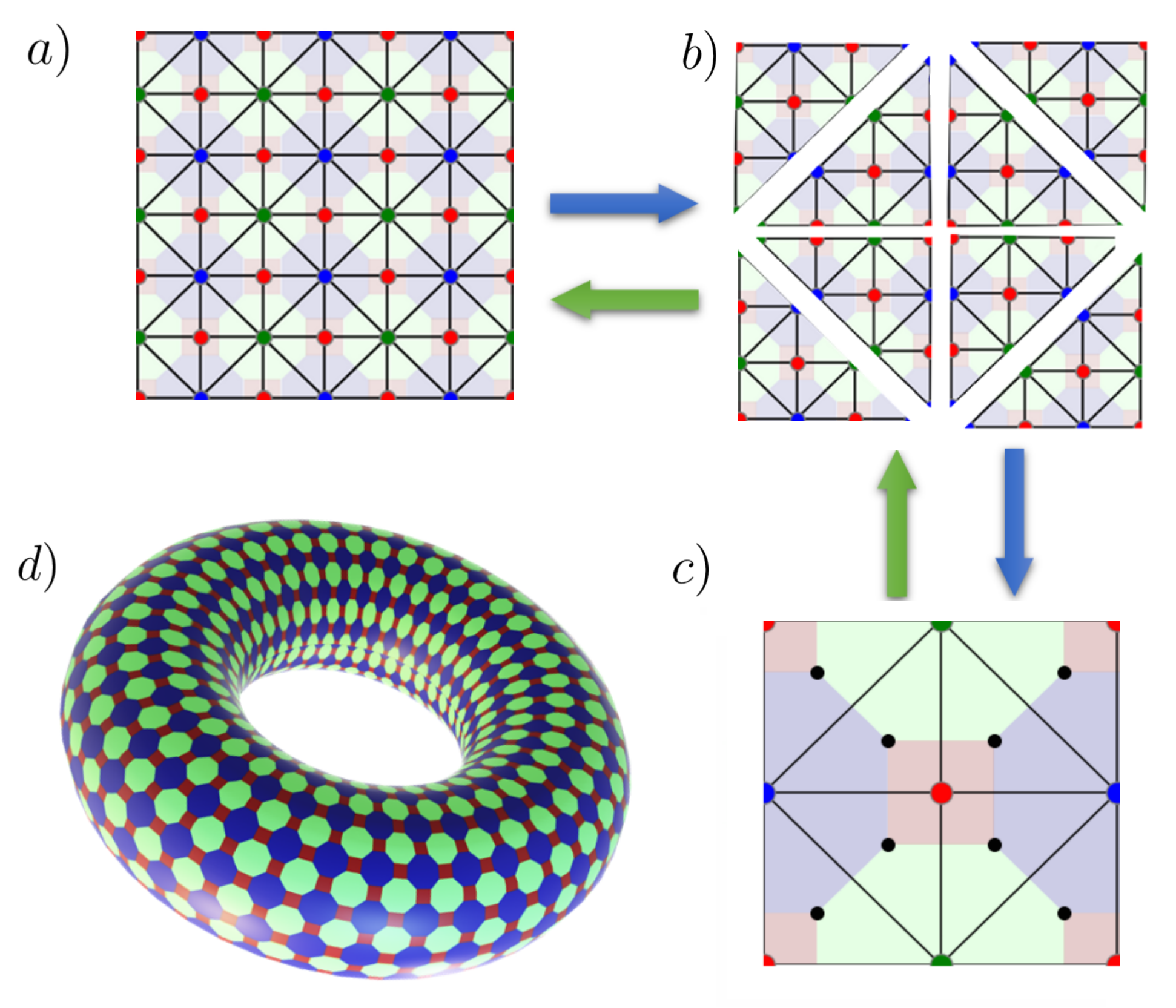}
\caption{ \textbf{Sketch of a rescaling step in the 4.8.8 color code lattice.}
Qubits are represented by triangular faces, and the coloured vertices represent the stabilizers.  The initial lattice \textbf{a)} can be split into multiple cells \textbf{b)}, which can be decoded locally. Each triangular cell can then be mapped into a single effective qubit in a rescaled lattice \textbf{c)}. This process of rescaling can be repeated until the final lattice is small enough to apply a brute force decoding. Corrections on the smaller lattices can then be backpropagated to the original lattice, finding the final correction.
\textbf{d)} In this work we consider a 4.8.8 color code lattice, where each qubit is involved in one 4-qubit stabilizer associated to a square plaquette and in two 8-qubit stabilizers on octagonal plaquettes, with periodic boundary conditions. The lattices considered host codes with parameters $[[n,k,d]] = [[8 \cdot 9^m,4,2\cdot 3^m]]$, where $n$ is the number of qubits in the lattice, $k$ the number of logical qubits, $d$ is the distance of the code, and the integer $m$ denotes the number of rescaling steps. The blue arrows represent the initial order of operations, in which the lattice is rescaled until the smallest system size is reached. Green arrows represent the backpropagation process, from the smallest rescaled lattice back to the original code.}
\label{fig4:488scheme}
\end{center}
\end{figure}

The paper is organized as follows. First, in Sec.~\ref{sec.background} we introduce central concepts about 2D color codes and the decoder proposed in~\cite{Sarvepalli_2012}, as well as some of the key ideas and concepts used in the algorithm. Then, in Sec.~\ref{sec.outlineDecoder} we explain the details about the decoder and the different steps: belief propagation, splitting of the stabilizers, and rescaling of the cells. In Sec.~\ref{sec.results}, we estimate the threshold for code capacity noise, obtained from Monte Carlo simulations for several lattice sizes. In Sec.~\ref{sec.conclusions}, we conclude and present some directions for future extensions of the decoder.

\section{Background\label{sec.background}}

\begin{figure}[ht!]
\begin{center}
\includegraphics[width=0.99\columnwidth]{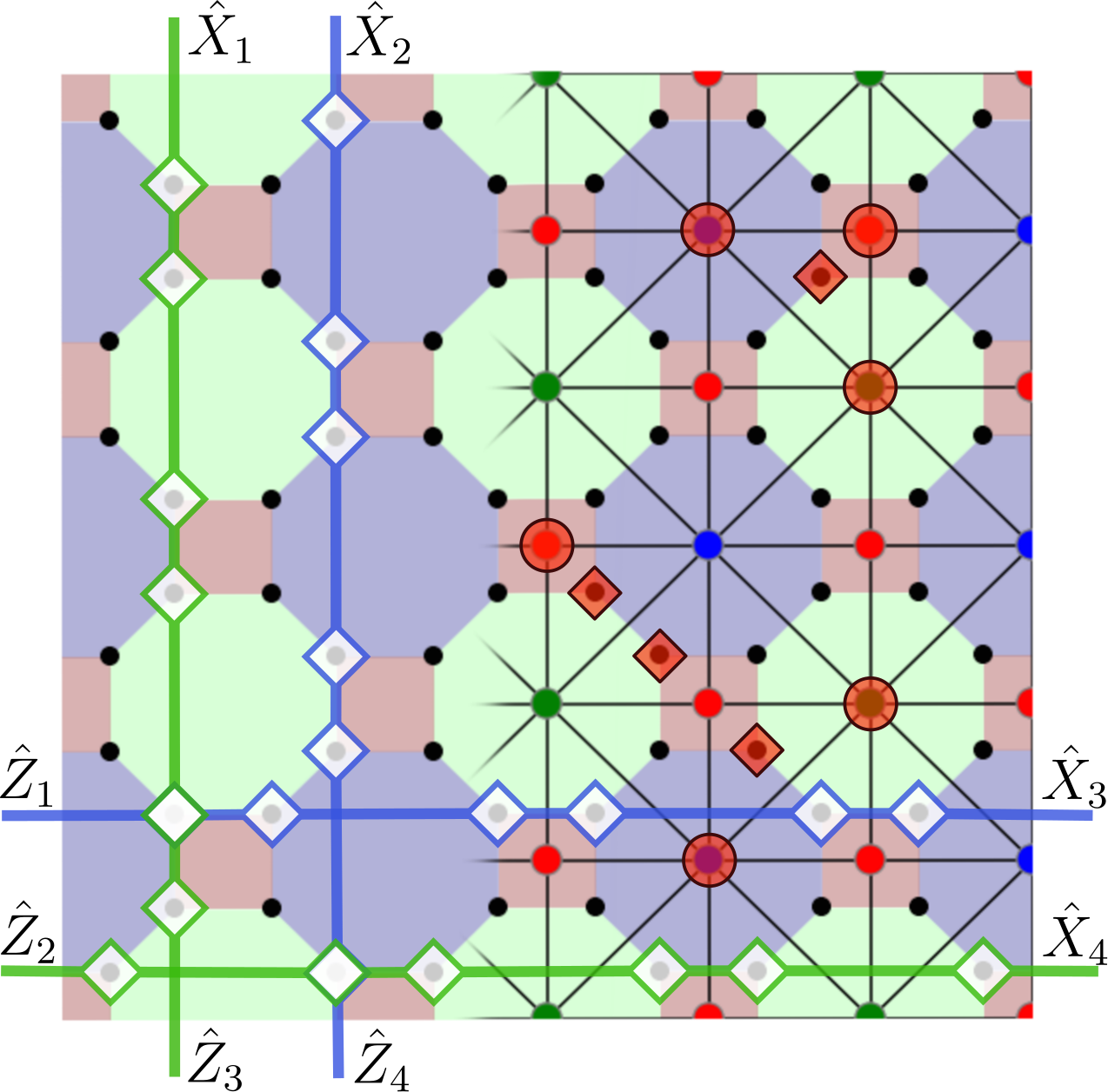}
\caption{ \textbf{Representation of the color code 4.8.8 lattice.} On the primal lattice (left), qubits are represented by black vertices, and the stabilizers $S_Z$ and $S_X$ are represented by the colored plaquettes, which apply a parity check over the qubits on the vertices. In the reciprocal lattice (right), the qubits are represented by triangles, and stabilizers are represented by colored vertices, which apply a parity check over the qubits for which the stabilizer is a vertex. The logical operators are strings of Pauli operators that extend over the torus in a non-trivial way. On the color code lattice on a torus, we have two independent logical operators for each non-trivial loop, i.e.~four logical qubits. The support of the logical operators $\hat{X}_i$ and $\hat{Z}_i$ is represented by the blue and green lines, which represent the 4 non-trivial loops on the toric color code lattice. To illustrate the effect of errors in the lattice, we display an example of four physical bit-flip errors (qubits are marked with red diamonds), and the corresponding stabilizer excitations (stabilizers are marked with red circles).}
\label{fig:lattice}
\end{center}
\end{figure}

Color codes are stabilizer codes~\cite{bombin2013introduction,colorCodes2006bombin,bombin2007colorcodes3d} defined on face-three-colorable trivalent graphs, i.e.~graphs where all vertices have degree three and faces can be colored with three colors such that neighboring faces never have the same color (see e.g.~Fig.~\ref{fig:lattice} for the case of the 4.8.8 lattice coloring). Qubits are identified with the vertices and each face $i$ of the graph defines two stabilizer generators $S_X^{(i)}$ and $S_Z^{(i)}$ involving all vertices in the boundary of the respective face~\cite{bombin2013introduction,colorCodes2006bombin}. These stabilizers involve purely $X$ or $Z$ Pauli operators, which render color codes part of the Calderbank-Shor-Steane (CSS) code family~\cite{calderbank1996css,steane-prl-77-793}. Note that stabilizers are guaranteed to commute since both faces as well as the boundary of faces on a trivalent three-colorable graph always contain an even number of vertices. The code space is defined as the simultaneous +1 eigenspace of all stabilizers. If we consider periodic boundary conditions, two independent qubits  can be encoded in each of the two non-trivial loops of the resulting torus (see Fig.~\ref{fig:lattice}), leading to a total of four logical qubits.  
Being part of the CSS stabilizer codes, it is possible to detect and correct phase-flip and bit-flip errors separately by using the syndrome from the $X$ and $Z$ stabilizers, respectively. Throughout this work we focus on independent bit- and phase-flip noise. One of the symmetries of the color code is that it is self-dual under exchanging $X$ and $Z$ stabilizers, which in particular implies that, in order obtain the code capacity threshold, we can focus purely on bit-flip errors. 

\section{The decoder algorithm}\label{sec.outlineDecoder}

\subsection{Introduction to the decoder's approach}

In this section, we introduce a qualitative description of the decoder algorithm. With this outline, we aim to frame the general picture of the decoder and the  basic underlying principle. In the following subsequent sections, we study the details of each particular step.

\begin{figure}[ht!]
\begin{center}
\includegraphics[width=0.7\columnwidth]{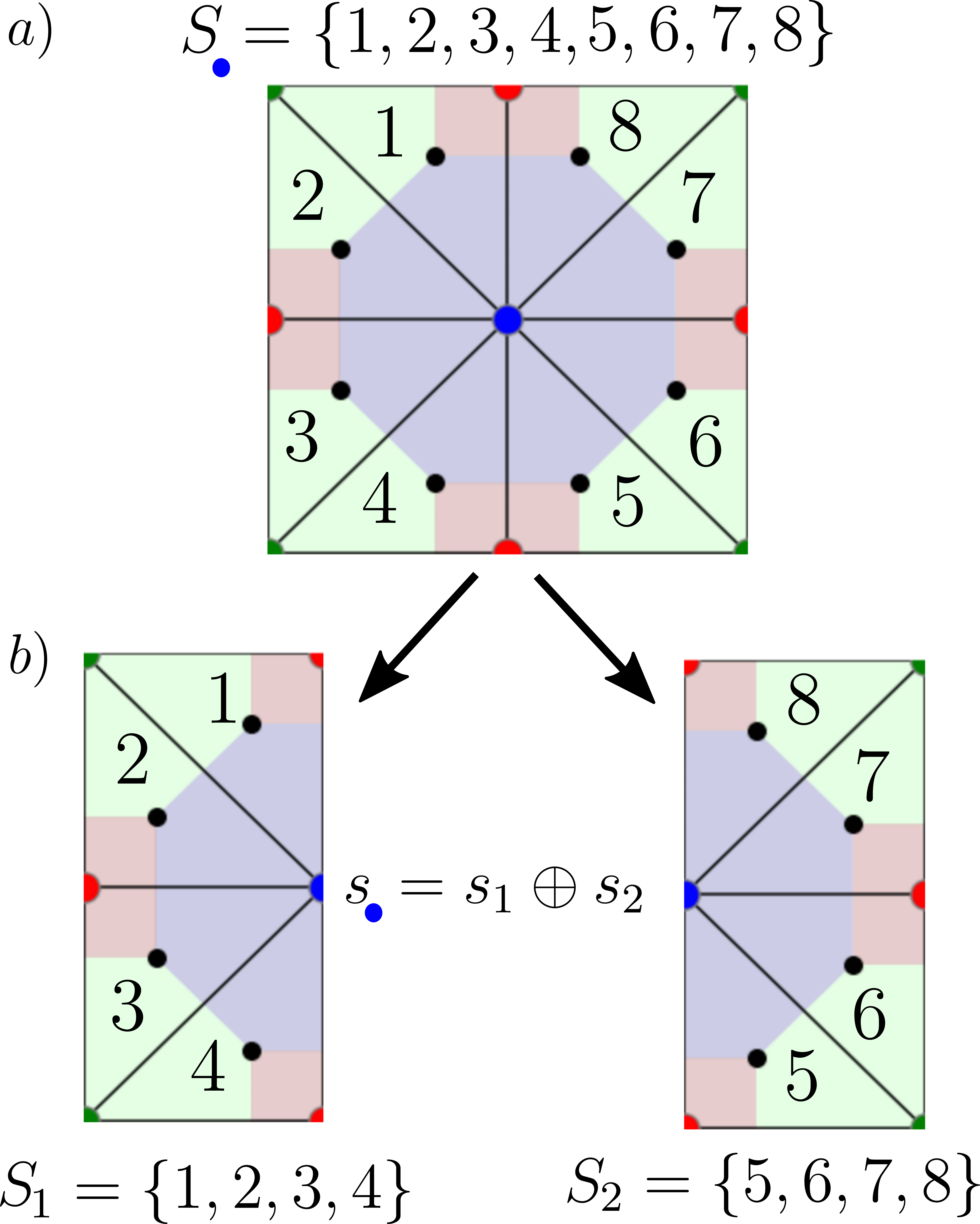}
\caption{ \textbf{Example of the splitting of a stabilizer.} \textbf{a)} The original stabilizer $S$, represented by the blue vertex, realizes a parity check on the qubits 1 to 8, both in the $Z$ and the $X$ basis. Qubits are represented by black dots inside the triangles.  \textbf{b)} We can split the stabilizer into the half stabilizers $S_1$ and $S_2$. Each half stabilizer represents the parity of its set of qubits $S_1=\{1,2,3,4\}$, $S_2=\{5,6,7,8\}$.  The binary sum $\oplus$ of the parities of both half stabilizers thus fulfills the condition $s= s_1\oplus s_2$. We call the choice of parity assignments of half stabilizers for a given stabilizer a \textit{splitting}. For notation, we use uppercase letters to refer to the stabilizer operators, and lowercase for their parity value, which is represented as a 0 for even parity and 1 for odd parity.}
\label{fig:splitting}
\end{center}
\end{figure}

The main idea behind the decoder is to split the code lattice into small cells. These cells can be decoded locally and the result can then be merged into the global decoding decision. A key problem to this decoding ansatz is that when trying to divide the lattice into cells, one invariably has to cut through some stabilizers that are shared between different cells. The solution to that problem is to split these stabilizers into two, which we will hence refer to as \textit{half stabilizers} in the following, so that one can decode each cell individually by using the local syndrome from the half stabilizer that applies to the cell (Fig.~\ref{fig:splitting}). Thus, the way in which each stabilizer is split between the cells determines the correction applied on the cells. After applying a local decoding on each cell, this cell can be treated as an individual effective qubit (two when using a square cell) of a now rescaled color code, where we have rescaled the lattice to a smaller version of itself. This process can then be repeated, ultimately leading to a lattice small enough so that a brute force decoding can be applied. This rescaling process is illustrated in Fig.~\ref{fig4:488scheme}(a)-(c) for the 2D color code on the square-octagon lattice.

During the algorithm, each qubit is assigned an error probability. This value can be initialized according to the error model, i.e.~in our case every qubit a priori has error probability $p$. This value is then updated using a belief propagation algorithm \cite{MacKay2003, Yedidia2003} to propagate information about the observed syndrome to the qubits. Using these error rates, the next step of the algorithm is to split the stabilizers that are shared between two cells into half stabilizers. In order to be consistent with the observed syndrome, e.g.~a stabilizer operator $S$ with parity $s=1$ (odd) can be split in two different ways: $(s_A,s_B) = (1,0)$ or $(s_A,s_B)=(0,1)$, so that the total parity of the sum of the half stabilizers correspond to the original value of the stabilizer $s = s_A \oplus s_B$. The choice of the splitting configuration determines the syndrome assigned to each individual cell. This syndrome is used by the decoder for a local cell to find a suitable correction within the cell. To find the best choice of splitting for the stabilizers (i.e.~the one that leads to the recovery operation for the most probable error), we apply a series of splitting updates, which assign a likelihood to each choice depending on the local information of the neighboring cells. We use a suitable convergence criterion, after which we then fix the configuration of splitting choices for the stabilizers, such that each cell can now be decoded using only the information from the local syndrome. Using the probabilities of the different error configurations compatible with that local syndrome, we can then rescale the cells to effective qubits and assign a new error probability to them. With this last step, we complete the rescaling cycle, which thereby leads to a smaller version of the lattice.

The challenge of defining a rescaling algorithm on the 4.8.8 lattice lies in identifying a valid choice for the form and size of the minimal cell. Due to the conditions that a cell must fulfill to be rescalable to an effective qubit, the minimal cell on the 4.8.8 lattice is more than twice times bigger and shares two stabilizers with each neighboring cell when compared to the simpler case of the 6.6.6 lattice, where neighboring cells share only a single stabilizer. This change requires the splitting algorithm that splits the stabilizers between cells to update splittings pairwise instead of individually.

\subsection{Decoder algorithm for the 4.8.8 lattice}

In the following we present the decoding algorithm in the form of high-level pseudo-code and discuss the main steps the decoder, which uses the measured syndrome and a prior estimate of the qubit error rate as input:\\

\begin{algorithm}
\caption{Rescaling decoder}\label{alg:three}
\DontPrintSemicolon

\SetInd{0.6mm}{4mm}
\KwData{Measured syndrome}
\KwResult{Recovery operation}
 \While{Lattice larger than minimum size}{
  Estimate the error probability of each qubit (1.)\;
  Split the code into cells (2.)\;
 Split the stabilizer values between the cells (3.)\;
 Apply a local decoder on each cell (4.)\;
 Rescale the cells to effective qubits (5.)\;
 }
  Apply a brute force decoder on the final lattice (6.)\;
 Back-propagate the errors to the original lattice (7.) \;
\end{algorithm}

\begin{figure}[ht!]
\begin{center}
\includegraphics[width=0.65\columnwidth]{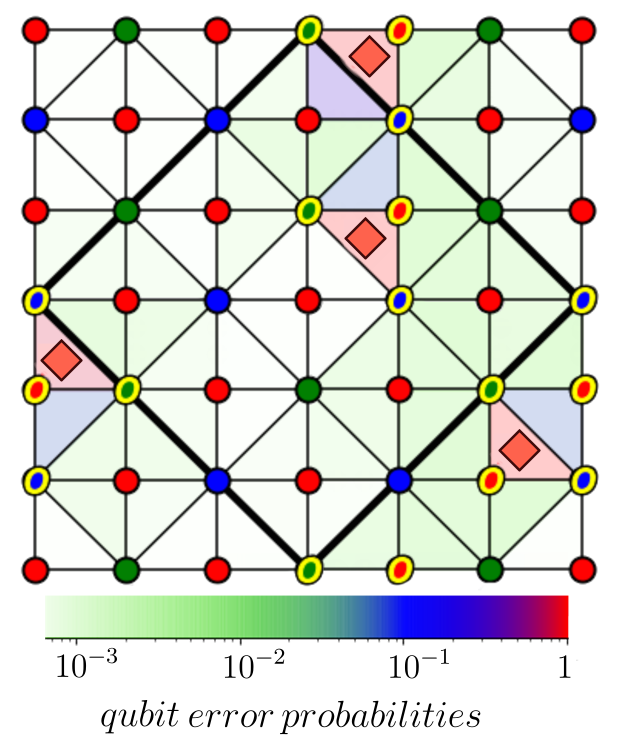}
\caption{ \textbf{Example of updated qubit error probabilities after belief propagation.}  We show a 72 qubit color code with periodic boundary conditions, where qubits are represented by the coloured triangles, and stabilizers are represented by the coloured circles in red, blue and green. The non-trivial stabilizers are represented with a yellow border, and the physical errors are represented by   red diamonds. During belief propagation, stabilizers and qubits share information locally, which leads to a refined estimate of the qubit error probabilities. In this example, the qubit error probabilities are represented by the coloring of the triangles, which correspond to the qubits, ranging from white to red as indicated in the color bar. The thicker black solid line has been drawn as a guide to the eye when comparing the figure with Fig.~\ref{fig4:splittinglattice}, after dividing the lattice into square cells.}
\label{fig4:bpcase}
\end{center}
\end{figure}

(1.) We estimate the error probability of the qubits using the information from the syndrome (Fig.~\ref{fig4:bpcase}). This is done in two steps. Firstly, we apply a belief propagation algorithm, through which stabilizers and qubits share information via message passing (see section~\ref{sec.beliefPropagation}). Secondly, we update the probabilities of the qubits around the corners of the cells depending on the parity of the stabilizer in the corner~\cite{Sarvepalli_2012}.\\

\begin{figure}[ht!]
\begin{center}
\includegraphics[width=0.9\columnwidth]{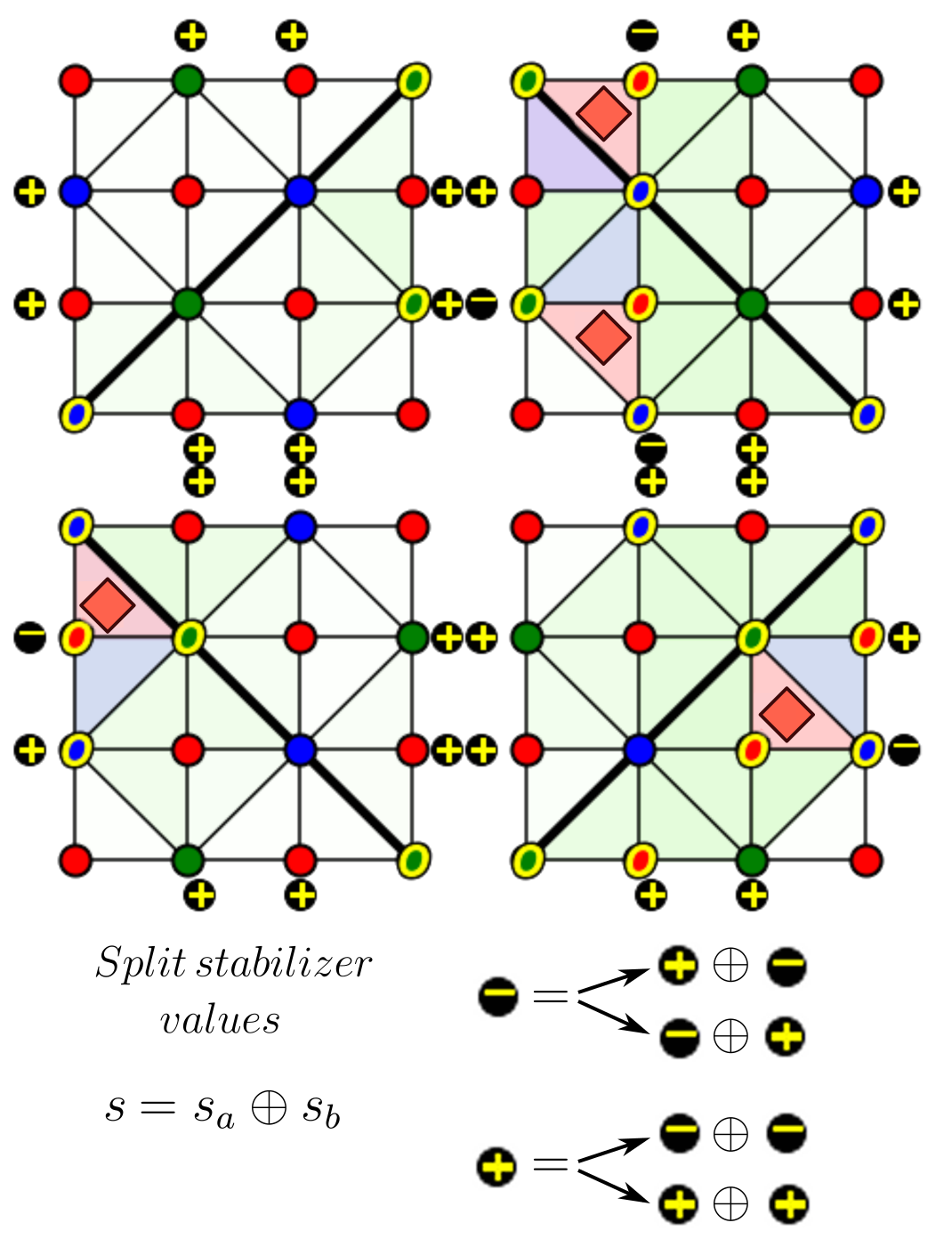}
\caption{ \textbf{Splitting of the lattice into cells.}  We show the same code as in Fig.~\ref{fig4:bpcase} after splitting the lattice into 4 different cells. The stabilizers at the boundary between cells need to be split into half stabilizers, as in Fig.~\ref{fig:splitting}. The parity of each half stabilizer $s_a$ and $s_b$ is represented as a $0$ for even parity and a $1$ for odd parity, and the binary sum $\oplus$ of the parity of both half stabilizers needs to equal the parity $s$ of the original stabilizer $S$. For each stabilizer, there are two alternative splittings into half stabilizers. The black solid line in the cells has been drawn to distinguish the two effective qubits that result from the rescaling of the cells. The physical errors in this particular example have been marked with red diamonds, and the excited stabilizers are marked with a yellow border.}
\label{fig4:splittinglattice}
\end{center}
\end{figure}

(2.) We subdivide the code into multiple cells. The stabilizers $S$ between two adjacent cells are split, while preserving the parity of the sum: $s_a \oplus s_b = s$. This allows for two ways of splitting a stabilizer (Fig.~\ref{fig4:splittinglattice}).\\

(3.) We find a configuration for the stabilizer splittings (Fig.~\ref{fig:splitting}) and assign a probability for each splitting choice (details will be provided in Sec.~\ref{sec.splittings}):

(3.a) We compute the initial splitting probabilities using the probabilities of the qubits involved.

\begin{figure}[ht!]
\begin{center}
\includegraphics[width=0.9\columnwidth]{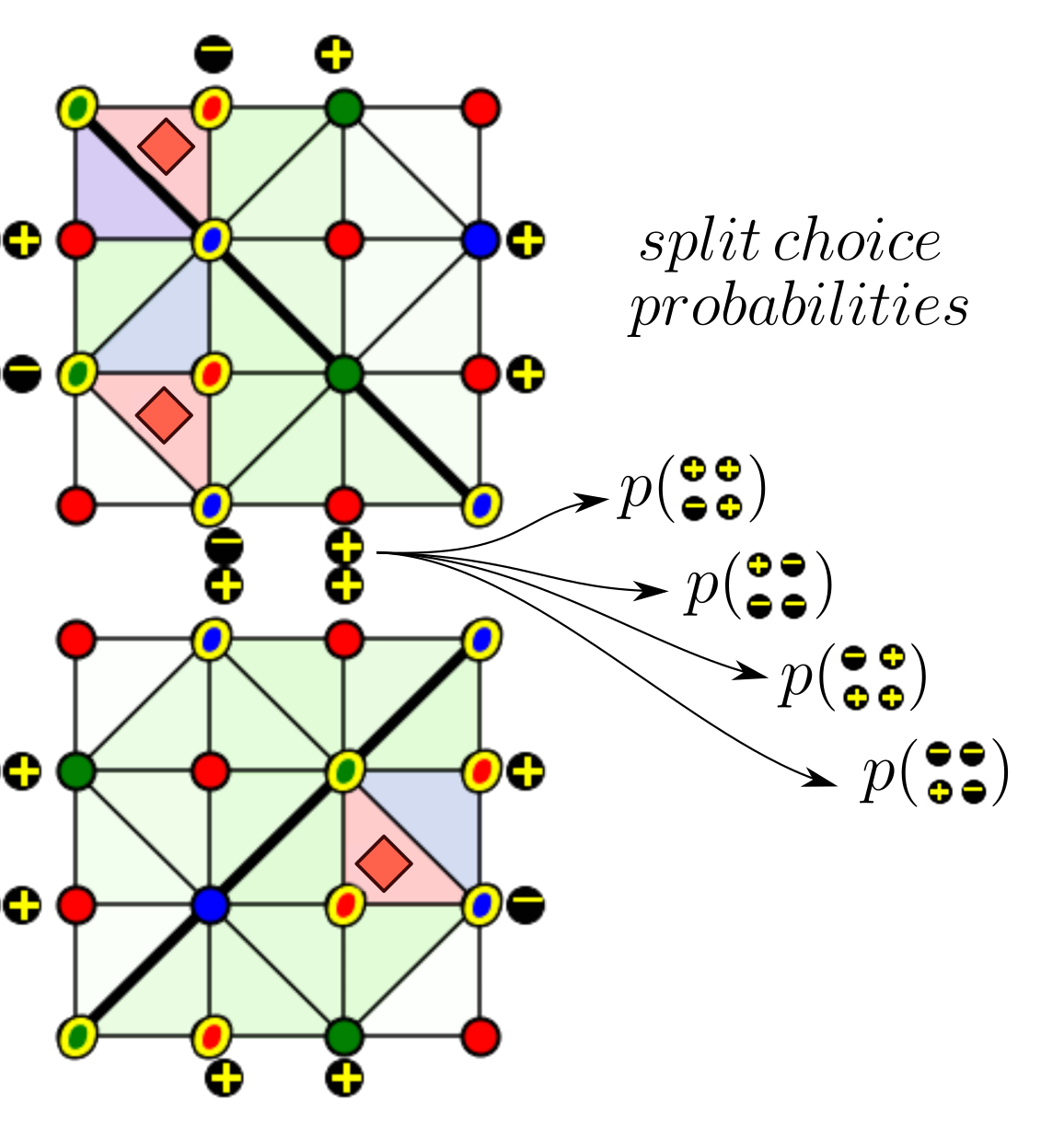}
\caption{\textbf{Sketch of the splitting of two stabilizers.} A cell shares two stabilizers with each of the neighboring cells. There are four different ways to split a pair of stabilizers in half stabilizers. During the splitting updates, we compute an estimate of the probability of each of the four splitting choices. The signs shown in the dark circles correspond to the parity of the half stabilizers. The two cells shown in the figure are part of the lattice shown in Fig.~\ref{fig4:bpcase}.}
\label{fig4:splitprob}
\end{center}
\end{figure}

(3.b) We update the probabilities of the different splitting choices (Fig.~\ref{fig4:splitprob}) using local information, as described in Sec.~\ref{sec.splittings}. The update is applied simultaneously on all splittings of the lattice. Several global update steps are used until convergence is reached for the split choice.\\

\begin{figure}[ht!]
\begin{center}
\includegraphics[width=0.6\columnwidth]{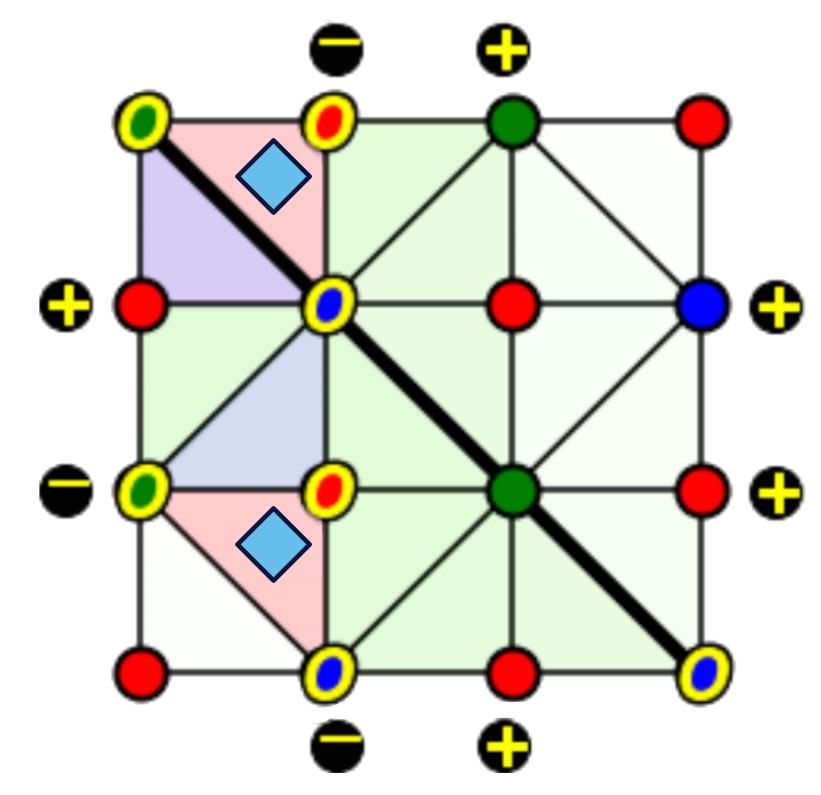}
\caption{ \textbf{Local decoding of cells.} A single cell can be decoded using the syndrome from the half stabilizers in the boundary and the parity of the bulk stabilizers. To find the correction, a brute force decoder is applied, which finds the most probable correction (the qubits in this suggested correction are marked by blue diamonds) using the estimates of the error probabilities of each qubit. The parity of the corner stabilizers is ignored during the decoding of the cell, but the parity of these stabilizers is updated depending on the corrections applied. In this particular example, the parity of the stabilizer in the upper left corner would be changed by the proposed correction, which will be taken into account in the rescaled lattice. The signs in the dark circles represent the parity of the half stabilizers in this cell. }
\label{fig4:cellDecoder}
\end{center}
\end{figure}

(4.) After fixing the half stabilizers, each cell has a local syndrome. Using a look-up table (a pre-computed list containing all possible errors compatible with the syndrome), we can decode the cells and find a correction (Fig.~\ref{fig4:cellDecoder}). At this step, we ignore the syndrome of the corners, as their parity will become the syndrome of the rescaled lattice. The parity of those stabilizers will thus be addressed in the decoding process applied to the subsequent rescaled lattices.\\

(5.) The cells can now be rescaled to effective qubits (Fig.~\ref{fig4:488cell}). 
We can compute the error probability of the rescaled qubits as the probability of applying a logical operator in the cell (see Sec.~\ref{sec.rescaling}).\\

(6.) We create a new code by rescaling each cell on the lattice to effective qubits. If the code is small enough, we can decode the lattice by finding the most probable error with a look-up table. Otherwise, we repeat steps 1-5 for the new code.\\
 
(7.) Once we have the corrections on the smallest lattices, we can back-propagate the corrections to the original lattice to obtain the final recovery operation.

After applying the recovery operation, we can check in our simulations if the combination of the error and our correction corresponds to the application of a logical operator, and thus the occurrence of a logical error, on any of the four encoded qubits in the lattice. This can be easily done by checking the parity of the qubits along each of the four operators drawn in Fig.~\ref{fig:lattice}, i.e.~for instance the parity of the number of bit-flips along the qubits in the support of $\hat{Z_i}$ determines if a logical $\hat{X_i}$ operator was applied.

It is important to notice that the smallest lattice size and the size of the unit cell determine the code size (total number of qubits) for which the decoder can be applied. In this work, we consider the minimum lattice size of 8 qubits. Thus, the lattice sizes for which the decoder can be applied depend on the number of rescaling steps $m$ as $[[n,k,d]] = [[8 \cdot 9^m,4,2\cdot 3^m]]$, where $n$ is the number of qubits in the lattice, $k$ the number of logical qubits and $d$ is the distance of the code. The number of qubits in the lattice begins with 8 qubits as the minimum lattice size (see Fig.~\ref{fig4:488scheme}), and each rescaling step introduces a factor of 9 in the number of qubits (see Fig.~\ref{fig4:488cell}). Similarly, the code distance $d$ of the smallest lattice size is 2, and each rescaling step increases the logical distance by a factor of 3 (see logical operators in Fig.~\ref{fig:lattice}). 

\subsection{Minimal cell}
In order to choose an appropriate cell, these are the conditions that need to be fulfilled (cf.~\cite{Sarvepalli_2012}):

\begin{enumerate}
    \item A logical operator can be defined for the cell.
    \item There exists a valid correction for every possible syndrome.
    \item The cell can map the entire code to a smaller version of itself.
\end{enumerate} 

A valid cell for which these conditions are fulfilled can be found by choosing a triangular cell for which the three corners are of different color.  

\begin{figure}[ht!]
\begin{center}
\includegraphics[width=0.85\columnwidth]{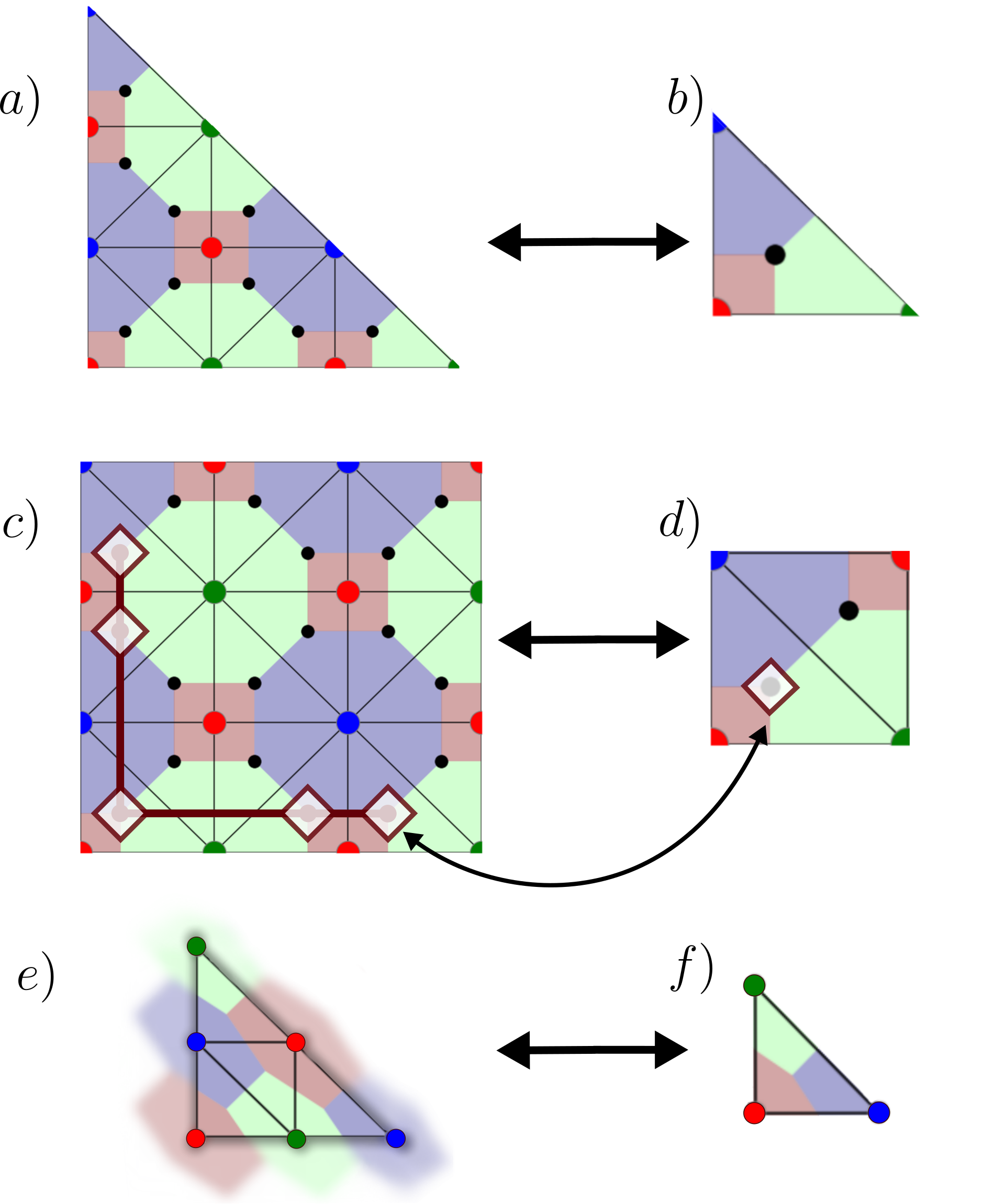}
\caption{\textbf{Rescaling cells in the 4.8.8 lattice.} \textbf{a)}~Minimal cell for the rescaling decoder in the 4.8.8 lattice. This 9-qubit cell can be mapped to a single effective qubit~\textbf{b)} during the rescaling of the lattice. 
\textbf{c)}~A square cell can be used during the decoder process to reduce the number of stabilizer splittings. This cell is mapped to two effective qubits, as shown in~\textbf{d)}. 
An error on the rescaled qubit can be back-propagated (curved double-arrow) to the original lattice by applying the effective logical operator of the cell: the qubits marked in \textbf{c)} with white squares represent the logical operator of the effective qubit marked in \textbf{d)}. By applying this logical operator, the parity of the corner stabilizers corresponding to the effective qubits is changed, while preserving the parity of the rest of the stabilizers in the cell. \textbf{e)}~For comparison, we show the minimal cell for the rescaling decoder in the hexagonal lattice~\cite{Sarvepalli_2012}, with only four qubits. This cell can be rescaled to an effective qubit~\textbf{f)}.  }
\label{fig4:488cell}
\end{center}
\end{figure}

The minimal cell that fulfills these conditions in the 4.8.8 lattice is the 9-qubit cell represented in Fig.~\ref{fig4:488cell}. The cell shares two split stabilizers with each neighboring cell, and an additional stabilizer is contained entirely within the cell. The syndrome for this cell contains seven stabilizer measurements (one red stabilizer inside the cell, and six half stabilizers on the boundaries), and four different corrections are possible for each possible syndrome due to the logical operator and the stabilizer contained within the cell. 

Using this cell, it is possible to map a 4.8.8 lattice to a smaller version of itself, reducing the number of qubits by a factor of 9 with each step. In Fig.~\ref{fig4:488scheme}, a single step of splitting the lattice and rescaling the cells is shown. In practice, it can be convenient to combine two triangular cells into a square cell consisting of two effective qubits (as shown in Fig.~\ref{fig4:cellDecoder}). By using square cells, we reduce the number of stabilizer splittings needed in each decoding step. This approach also proved to achieve better correction capabilities in the hexagonal lattice~\cite{Sarvepalli_2012}. In this work, we study the performance of the decoder using these square cells. 

\subsection{Belief propagation}\label{sec.beliefPropagation}
In general, belief propagation (BP) is a method to compute or approximate marginal probabilities of multivariate probability distributions~\cite{Pearl1982}. BP is formulated in a graph-theoretical setting, where the multivariate distribution is represented as a factor graph, which is a bipartite graph with vertices representing a) the variables and b) the factors such that edges between the two indicate a functional dependence of the latter on the former. In an error correction setting, the role of variables is taken by the qubits and the role of factors by the parity checks, i.e.~the stabilizer generators and the multivariate probability distribution is the error model (e.g.~independent bit-flip noise, depolarizing noise, etc.). The task of marginalization, i.e.~of summing over all variables except the one we are interested in, is in general exponentially hard, since there are exponentially many configurations to sum over. However, the ingenuity of BP (or more precisely the sum-product message passing algorithm~\cite{MacKay2003}) lies in streamlining the summations into subtasks, called messages, which are computed locally and then passed forward on the edges of the factor graph. If the structure of the factor graph is tree-like (i.e.~contains no loops), this renders the BP algorithm exact, i.e.~it computes the exact marginals on all variables in a number of steps proportional to the depth of the tree. While most graphs are not tree-like, it turns out that in practice BP can still yield good approximations to the marginals as long as the underlying graph does not contain too many loops - as a decoder for classical so-called low-density parity check (LDPC) codes it performs close to optimal~\cite{MacKay1996}.

In general, as a standalone decoder, BP is known to fail when trying to decode topological QEC codes, which is attributed to the highly degenerate nature of topological codes~\cite{Poulin2008}: degeneracy refers to the fact that one syndrome can be corrected in many distinct ways that are nevertheless logically equivalent. Here, this problem does not arise since we abort BP iterations before running into loops. For the decoder investigated here, BP serves as a means to compute estimates of the qubit error probabilities from the stabilizer measurement outcomes, which then acts as input to the remaining processing steps.

The main idea behind the algorithm is the following: 
\begin{itemize}
    \item At the beginning, each qubit has an estimate of its error probability. Each syndrome has a parity value.
    \item Each qubit sends a message to the neighboring stabilizers with information about its error probability.
    \item With the information from the error probabilities of the qubits involved in a parity measurement, each stabilizer can send an updated estimate of the error probability for each qubit. For example, if the parity is even, this means that an error in a qubit implies an odd parity in the rest of the qubits involved in the stabilizer. Therefore, the stabilizer sends a message to each qubit with information on the probability of \textit{the rest} of the qubits having an odd number of errors. Similarly, if the parity is odd, the message will contain information about the probability of an even number of events in the remaining qubits.
    \item With the updated information from the stabilizers, the qubits send an updated message to the neighboring stabilizers, repeating the process. With each cycle of message passing, the information spreads through the code.
    \item After a given number of iterations, the algorithm stops and utilizes the information from the messages to update the estimate of the error probability of the qubits (see Fig.~\ref{fig4:bpcase}).

\end{itemize}
Now, let us get into the details of the mathematics behind the algorithm. The building block of belief propagation is Bayes theorem, applied to update the probability of an error in a qubit with the information from the syndrome:

\begin{equation}
    p(q_i=1|\{s\}) = \frac{p(q_i=1)}{p(\{s\})} \cdot p(\{s\}|q_i=1)
\end{equation}
\begin{equation}
    p(q_i=0|\{s\}) = \frac{p(q_i=0)}{p(\{s\})} \cdot p(\{s\}|q_i=0)
\end{equation}
where  $p(q_i=1|\{s\})$ is the probability of an error on qubit $q_i$ given the syndrome $\{s\}$, $p(q_i=1)$ is the prior, or the previous information on the qubit error rate, $p(\{s\}|q_i)$ is the probability of the given syndrome assuming an error on qubit $q_i$ (usually called the likelihood ratio) and $p(\{s\})$ is the probability of the syndrome event. This last term is effectively an unknown normalization factor, that would be hard to compute. However, we can cancel that term if we compute instead the quotient of both quantities. Thus, we can write:

\begin{align}
    \frac{p(q_i=0|\{s\})}{p(q_i=1|\{s\})} =& \frac{p(q_i=0)}{p(q_i=1)} \cdot \frac{p(\{s\}|q_i=0)}{p(\{s\}|q_i=1)}\\ \nonumber
    =& \frac{1-p_i}{p_i}  \prod_j \frac{p(s_j|q_i=0)}{p(s_j|q_i=1)},
\end{align}

where the product in $j$ compiles information for all the parity  $s_j$  of the stabilizers affecting the qubit. This quotient will correspond to the information sent from each stabilizer $j$ to the qubit, and can be written, depending on the parity of the stabilizer, as
\begin{equation}\label{eq4:messagesq1}
    M_{s=0 \rightarrow q}= \frac{p(s=0|q_i=0)}{p(s=0|q_i=1)}=\frac{p(even)}{p(odd)},
\end{equation}
\begin{equation}\label{eq4:messagesq2}
    M_{s=1 \rightarrow q}=\frac{p(s=1|q_i=0)}{p(s=1|q_i=1)}=\frac{p(odd)}{p(even)}.
\end{equation}
Here, $p(even)$ and $p(odd)$ correspond to the probability of an even/odd number of error events happening on the remaining qubits involved in the parity check. 
Therefore, we can compute the messages from the stabilizers by computing the probability of an even/odd number of error events happening on the remaining qubits. This probability can be written using a simple formula to find the probability of an even/odd number of error events, given the probabilities $p_i$ of each individual event:

\begin{align}\label{eq.evenodd}
p(even| \{p_i\})&=\frac{1}{2}+ \frac{1}{2} \prod_i (1-2 p_i),\\
p(odd| \{p_i\})&=\frac{1}{2}- \frac{1}{2} \prod_i (1-2 p_i). \label{eq.evenodd2}
\end{align}

Then, the messages from the qubits to the stabilizers are updated using the information from the stabilizers. The messages to the stabilizers in the next cycle are:

\begin{equation}\label{eq4:messageqs}
    M_{q\rightarrow s_i} = \frac{1-p}{p} \prod_{\substack{j \in N(q) \\ j \neq i}} M_{s_j \rightarrow q},
\end{equation}
where $N(q)$ refers to the three stabilizers in the neighborhood of qubit $q$.
In this way, the information from the qubits and stabilizer spreads through the code, leading to an improved estimation of the error probabilities for the qubits. However, in the numerical simulations, the direct use of Eqs.~(\ref{eq4:messagesq1}),~(\ref{eq4:messagesq2}) and~(\ref{eq4:messageqs}) leads to numerical problems, when the error probabilities of the qubits assume values close to zero. To solve this problem, it is useful to work with the log-likelihood ratio. With that formulation, the initial message from qubits to stabilizers is

\begin{equation}
M_{q\rightarrow s}^{(0)}  =  \log \frac{1-p_q}{p_q}.
\end{equation}
The equation for the messages from stabilizers to qubits can be written as:
\begin{equation}
    M_{s_i \rightarrow q_j} = (1-2\cdot s_i)\, 2\,\tanh^{-1} \left[ \prod_{\substack{k \in N(s_i)\\ k\neq j}} \tanh ( M_{q_k \rightarrow s_i}/2) \right]
\end{equation}
where $s_i$ represents the parity of the stabilizer measurement. Similarly, the equation for messages from qubits to stabilizers is simplified as:
    \begin{equation}
        M_{q_i \rightarrow s_j} = \sum_{\substack{s_k \in N(q_i)\\ k \neq j }}M_{s_k \rightarrow q_i} + \log \tfrac{1-p_i}{p_i}.
    \end{equation}
    
    After the last iteration of message passing, we can obtain the updated estimate of the error probability. The equation for the final estimate of the error probability can be written as:
    
    \begin{equation}
        p_i^{updated} = \left[1 + \exp\left(\log \tfrac{1-p_i}{p_i} + \sum_j M_{s_j \rightarrow q_i} \right) \right]^{-1}.
    \end{equation}

\subsection{Corner Updates}
\label{sec.cornerupdates}

During the decoding process, the splitting updates and the cell decoder ignore the syndrome of the stabilizers located in the corners of the cells. This can be problematic for some error cases, as the decoder would not be able to distinguish cases that differ only on the syndrome of the corner stabilizers. For this reason, it is useful to use the information from the syndrome in the corners of the cells to modify the error probability of the qubits near the corners. 

Given the parity $s_i$ of a given corner stabilizer, we need the parity of the qubits in the cell plus the parity of the qubits outside to be equal to $s_i$. Thus, for a given qubit $j$ in the cell with a prior estimate $p_j$, we can compute the updated error probability of an error on qubit $j$ given the parity of the corner stabilizer $p(j|s_i,p_e)$ as:

\begin{eqnarray}
    p(j|s_i =0) =& \frac{p_j \cdot p_e(odd)}{p_j \cdot p_e(odd)+(1-p_j)p_e(even)}, \\
    p(j|s_i =1) =& \frac{p_j \cdot p_e(even)}{p_j \cdot p_e(even)+(1-p_j)p_e(odd)}, 
\end{eqnarray}
where $p_e$ corresponds to the probability of the qubits outside the cell to have a total parity that is even/odd. 
We can use the Eqs.~(\ref{eq.evenodd}) and (\ref{eq.evenodd2}) to compute the probability of $n$ qubits having a total even or odd parity.

\subsection{Splitting of the stabilizers}\label{sec.splittings}

The basic cells on the 4.8.8 lattice share two splittings with each neighboring cell. This leads to a new phenomenon compared to the simpler case of the 6.6.6 lattice, as the probability of splitting of each stabilizer now depends on the splitting configuration of the other splitting shared with that cell. Therefore, we need to consider the joint probabilities, which take into account simultaneously the probability of a splitting configuration of the two stabilizers $s_1$ and $s_2$ shared between the cells. 

This complication in the way we store and compute the probabilities of splitting configurations will affect most of the other steps of the algorithm, as splittings need to be considered in pairs, taking into account the joint probabilities:
\begin{enumerate}
    \item The first estimate for the joint probabilities is obtained from the probabilities of the qubits involved in each stabilizer having an even/odd number of errors. This does not take into account the rest of the splittings in the cell, only the probabilities of the qubits involved in the splitting.
    \item During the splitting updates, we will update the joint probabilities for the splitting configurations. Each step will update our estimate for the probabilities of each splitting using the information from the two cells involved.
    \item In the rescaling step, the probability of error in the rescaled qubit involves the probabilities of the different splitting configurations. This means that the error probability of the rescaled qubit will take into account the uncertainty in the splitting choice.
\end{enumerate}

Now, let us get into the details of the equations needed to compute the updated probabilities for the splittings. The first equation we need to consider corresponds to the probability of a given error configuration. Assuming our estimate of the error probabilities of each individual qubit, we can compute the probability of an error configuration $C = \{e_0,e_1,...e_{n_q-1}\}$ (where $e_i=0,1$ represents no error or an error on qubit $i$, and $n_q$ is the number of qubits in a cell) by adding a factor of $p_i$ for each qubit with an error, and a factor of $1-p_i$ for each qubit without an error,
\begin{equation}
    p(C) = \prod_{i=0}^{n_q-1} p_i^{e_i} (1-p_i)^{1-e_i},
\end{equation}
where we assumed no correlation between the error probability $p_i$ of each qubit. During the rescaling process, the use of square cells with two qubits will give us access to the joint probabilities of each qubit pair that belongs to the same cell. We can use this additional information about the correlations between different qubits to modify this equation as:
\begin{equation}
    p(C) = \prod_{i=0}^{n_q/2-1} p (e_i,e_{i+1}),
\end{equation}
where $p(e_i,e_{i+1})$ corresponds to the element of the joint probabilities corresponding to our prior knowledge from the error probabilities of qubits $i$ and $i+1$. 

Using the probability of a given error configuration, we can compute the probability of all error configurations that are compatible with a given syndrome as the sum of $p(C)$ over all configurations $C$ compatible with the syndrome $\{s\} = \{s_0,s_1...s_{n_S-1}\}$ ($s_i=0,1$ represents the even or odd parity of the stabilizer or half stabilizer $i$):
\begin{equation}\label{eq4:sumAllConfigurations488}
    p(e|\{s\}) = \sum_C p(C).
\end{equation}
For the 18-qubit square cell that we use for the 4.8.8 color code, each syndrome includes eight half stabilizers (two on each side of the square) and four additional stabilizer measurements corresponding to the bulk stabilizers inside the cell. Here, as stated earlier, the parity of the corners is ignored, as the parity of the corner stabilizers will be solved in the following rescaling steps. For the square cell, there are two logical operators that can be defined, one for each of the logical qubits to which the cell will be rescaled. This means that there will be $2^2$ possible classes of configurations compatible with any syndrome in the cell. In addition to that, all possible product combinations with the four bulk stabilizers lead to equivalent error configurations. This leads to a total of $2^{2+4}=64$ possible configurations over which is to be summed according to Eq.~(\ref{eq4:sumAllConfigurations488}).

 While using all 64 configurations would lead to more accurate estimations, this also involves a high constant-factor overhead in the computational time required by the decoder. Thus, to improve the performance of the algorithm with regard to computing time, we  approximate Eq.~(\ref{eq4:sumAllConfigurations488}) by considering only one configuration, using the cell's look-up table to find the most likely configuration. With this approximation, we  effectively reduce the computational overhead by a factor of 64, while keeping one of the terms of highest weight in Eq.~(\ref{eq4:sumAllConfigurations488}). In our simulations, we found no significant change in the correction capabilities of the decoder for system sizes larger than 72 qubits. Therefore, we used this approximation for the simulations shown in Sec.~\ref{sec.results}.

\begin{figure}[ht!]
\begin{center}
\includegraphics[width=0.9\columnwidth]{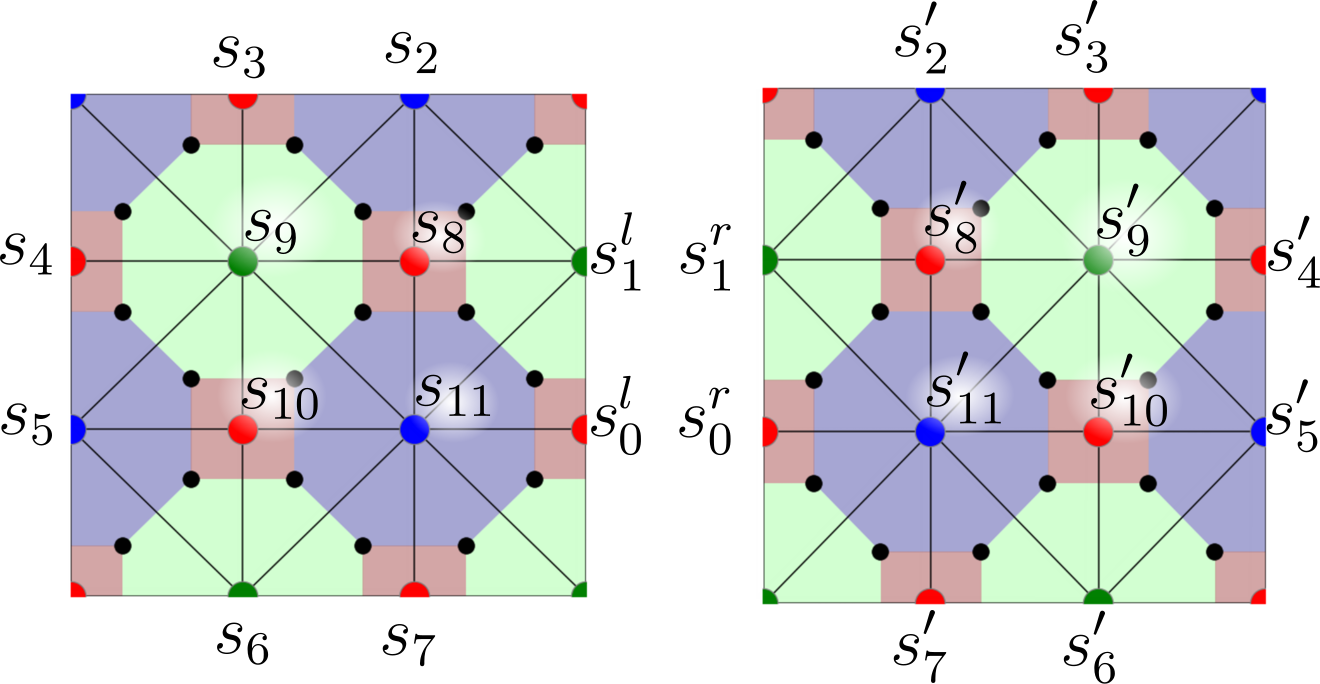}
\caption{ \textbf{Notation of stabilizer splitting.} Example of a pair of splittings between two neighboring cells in the 4.8.8 lattice. The splittings 0 and 1 are being updated, and the super index indicates if we refer to the left or right half stabilizer. The parity of the (half) stabilizers on the left cell are represented as $s_i$, while the parity of the (half) stabilizers on the right cell are notated with $s_i'$.}
\label{fig4:splitCellRef488}
\end{center}
\end{figure}

Once we can compute the probability of all possible configurations compatible with a given syndrome, we can define the probability of a given half splitting $s_i^l$ given a certain fixed syndrome on the rest of splittings (where the superscript $l$ corresponds to the cell on the left, see Fig.~\ref{fig4:splitCellRef488}). This probability corresponds to the fraction of possible configurations compatible with the splitting choice, compared with the total probability of the possible configurations for all splitting choices,
\begin{equation}
    p(s^l_i|\{s\}) = \frac{p(e|s_i,\{s\})}{p(e|s_i=0,\{s\}) + p(e|s_i=1,\{s\})}.
\end{equation}

Seeing that every cell shares two stabilizers with each of its neighbors, the splitting choice for these two stabilizers is not independent. Therefore, we need to consider the joint probabilities, where we consider each of the combined splitting choices for the two splittings shared between two cells. Therefore, the probability of a splitting choice for the stabilizers $s_0^l$ and $s_1^l$ given a fixed choice on the rest of splittings (which we write as $\{s\}$ for simplicity) can be written as

\begin{equation}\label{eq4:probSplitWithinCell488}
p(s_0^l,s_1^l|\{s\}) = \frac{p(e|s^l_0,s^l_1,\{s\})}{\sum_{i,j=0}^1p(e|i,j,\{s\})}.
\end{equation}

Note that this expression is equivalent for all splitting pairs, and we only wrote it explicitly for the first two splittings to simplify the notation. Since the rest of the splitting choices are not fixed, we can compute the estimate for the half splitting within a cell by combining the information from all splitting choices. For this, we can use the information of the joint probabilities $p(s_k,s_{k+1})$, which corresponds to our current estimate of the probability of the half splitting $s_k,s_{k+1}$ (e.g.~the probability of the splitting choice could  be $p(s_2= 1,s_3= 1)$). The estimate for the probability of a half splitting configuration within the cell can then be computed as the sum of Eq.~(\ref{eq4:probSplitWithinCell488}) for each splitting configuration of the other splittings involved in the cell

\begin{equation}
    p(s^l_0,s^l_{1}) = \sum_{\{s\}_k} p(s^l_0,s_1^l|\{s\}_k)\prod_{k=1}^{3} p(s_{2k},\,s_{2k+1}). 
\end{equation}

Finally, we want to find the probability of a given splitting choice. This necessarily involves the configurations in both the left and right cells, and the value of the parity $v_i$ of the stabilizers that we are measuring. Combining the information from both cells and ensuring the consistency condition $s_i^l \oplus s_i^r = v_i $, we can obtain the next estimate for the probability of a given splitting choice as

\begin{equation}
    p_{split}(s^l_0,s^l_1) = \frac{ p (s_0^l, s_1^l) p'(s_0^r =s_0^l \oplus v_0 , s_1^r=s_1^l \oplus  v_1)}{\sum_{i,j=0}^1 p(i,j)p'(i\oplus v_0,j\oplus v_1)}, 
\end{equation}
where $\oplus$ represents a binary sum, and $\bar{v_i} = 1\oplus v_i$. Using these equations, we can update our estimate for each of the splitting choices.

During the splitting algorithm, we apply global updates to the probabilities of the splitting choices by updating simultaneously the splitting probabilities of all splitting pairs in the code. All of the updates for the estimates of the splitting probabilities depend on the estimates from the previous step, which are not overwritten until all new estimates have been computed. 

\begin{figure}[ht!]
\begin{center}
\includegraphics[width=0.99\columnwidth]{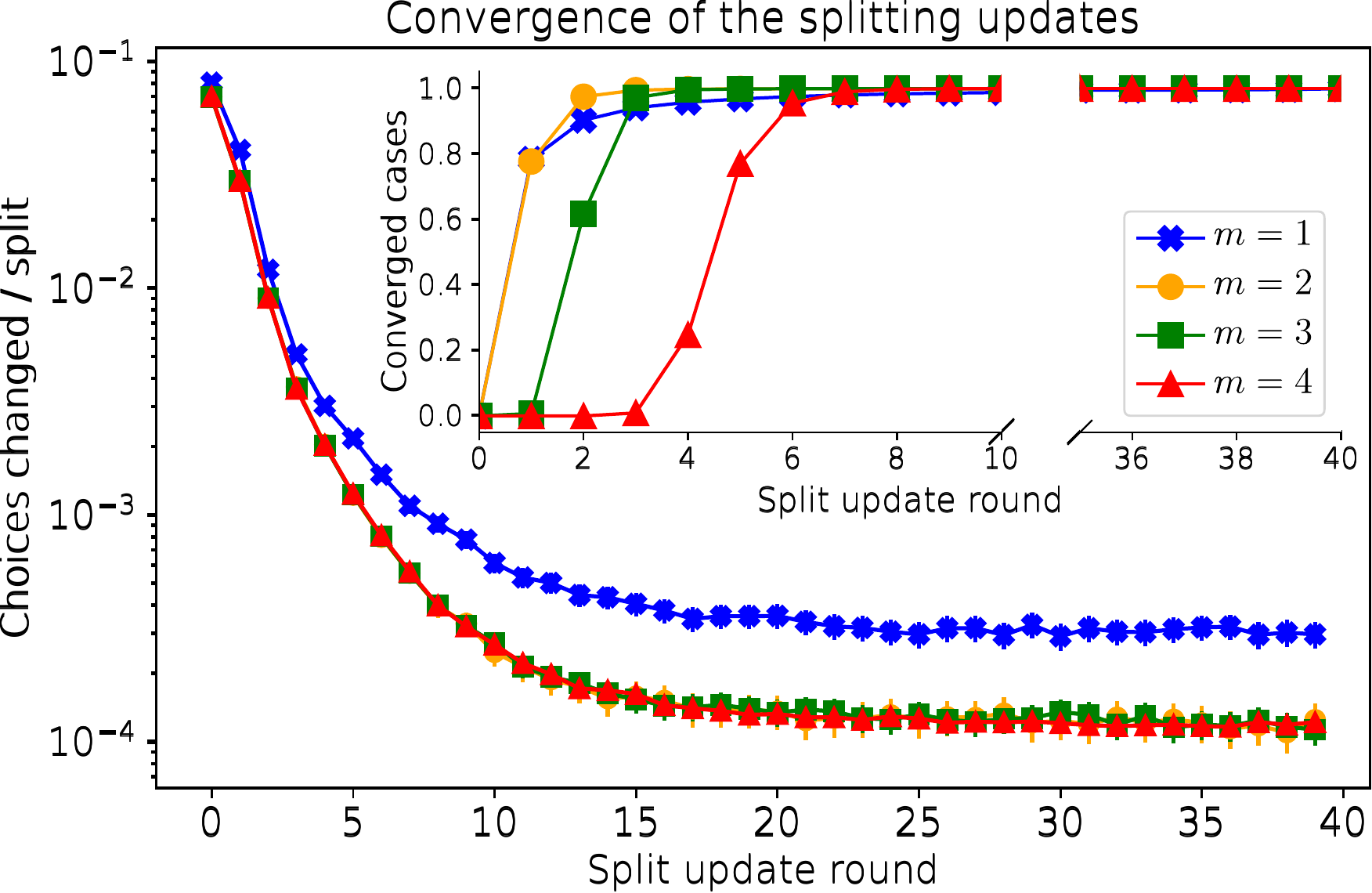}
\caption{\textbf{Convergence of the splitting updates.} We show the statistics of the performance of the splitting updates. In the main plot, we show the average fraction of splittings that have changed the split choice on each split update round. In the inset plot, we show the cumulative fraction of cases that have converged by each split update round. We define a case to have converged at round $r$ if that was the last round with more than $3m$ changes in the splitting choice. For both plots, the lines connecting numerical data points have been drawn as a guide to the eye.}
\label{fig4:splitconvergence}
\end{center}
\end{figure}

Although the number of global updates required for convergence can vary between runs, we empirically find that the average number of updates does not scale significantly with the lattice size, with less than 15 update rounds required. We test the convergence by measuring the average number of changes in the splitting choice per splitting and per round. We find that this ratio does not increase with the system size (Fig.~\ref{fig4:splitconvergence}). Furthermore, we estimate the fraction of cases that have converged after $n$ split update rounds. For a given error case, we define the number of rounds until convergence as the last update round with less than $3m$ changes in the split choice (with $m$ being the number of rescaling steps required for that lattice size).

\subsection{ Rescaling of the cells}\label{sec.rescaling}

After a splitting configuration is chosen, a correction can be found within each of the cells in the lattice. The next step in the decoder is to map each square cell to a pair of qubits. The error probability of each qubit on the rescaled lattice also needs to undergo a rescaling process. In this section, we discuss the details on how to compute what is the resulting error probability for the rescaled qubits.

The main idea to understand in the rescaling of the qubit error probability is the fact that an error on a qubit in the rescaled lattice corresponds to the application of a logical operator on the qubit of the original lattice. If the original correction in the cell is $C$, we can write a first estimate of the error probability of the rescaled qubit given the splitting choice $\sigma$ as:

\begin{equation}\label{eq4:conditionalRescaling488}
    p(L|\sigma)= \frac{ \sum_{\{S_b\}} p(C + L + \{S_b\})}{ \sum_{\{S_b\}} p(C+\{S_b\}) + p(C + L + \{S_b\})},
\end{equation}
where the sum over $\{S_b\}$ represents all possible combinations of the bulk stabilizers, $L$ is the logical operator and $p(E)$ is the probability of a given error configuration $E$. 

For this estimate, we assumed that the splitting choice from the previous step is correct. However, from the previous splitting step we know that we have an uncertainty in the splitting choice. In addition, we have an estimate of the probability of each splitting choice. Thus, we can include this information about the other splitting choices, weighted by the probability of each splitting choice, in the equation for the probability of an error in the rescaled qubit.

In order to include the information from alternative splitting configurations, we first need to understand what this probability of an alternative splitting means, and how to relate it to the correction $C$ that we applied on the cell during the previous step.  In particular, we need to find an expression of $p(L|\Tilde{\sigma}_k)$ for the alternative splittings $\Tilde{\sigma}_k$.

For the decoder of the 6.6.6 lattice~\cite{Sarvepalli_2012}, the key idea to find this expression is that, by applying the stabilizer of a given splitting, we can effectively change the choice of that splitting, as there is an odd number of qubits from that stabilizer on each cell. By applying the half stabilizer corresponding to the splitting, we could relate the different corrections that correspond to each splitting choice, thus finding an expression for $p(L|\Tilde{\sigma}_k)$. 

\begin{figure}[ht!]
\begin{center}
\includegraphics[width=0.99\columnwidth]{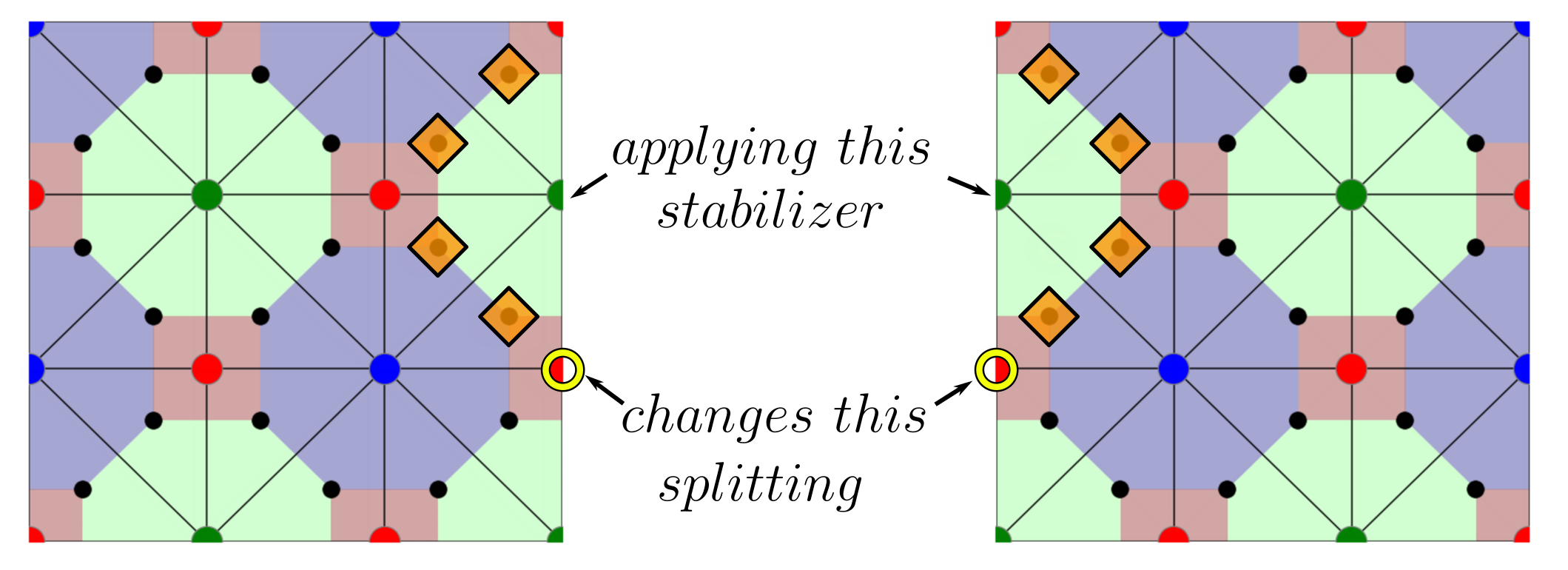}
\caption{Each cell shares two stabilizers with each of the neighboring cells, which need to be split. These two splittings form a pair of splittings. By applying the half stabilizer of one of the stabilizers, the resulting correction corresponds to a change in the splitting choice of the other splitting in the pair. In the figure, we can see one example in which, by applying the half stabilizer of one of the splittings, we change the parity of the neighbor splitting and obtain a valid correction for the new syndrome. The qubits affected by applying the stabilizers are marked with orange diamonds, and the stabilizer for which the splitting choice is changed is marked in yellow.}
\label{fig4:rescalingEquivalence}
\end{center}
\end{figure}

In contrast, on the 4.8.8 lattice, the support of the half stabilizers on each cell consists of an even number of qubits. This means that by applying the stabilizer, we do not change the value of the splitting choice for the splitting corresponding to that stabilizer. However, for the 4.8.8 color code, splittings come in pairs, as each cell shares two stabilizers. By applying the stabilizer of one of these two splittings, we can effectively change between the two choices of the neighbor splitting, and thus relate the corrections corresponding to both splitting choices. An example of this equivalence is shown in Fig.~\ref{fig4:rescalingEquivalence}.

Following this rule, we can systematically find the probability $p(L|\sigma_k)$ for each of the $2^8$ possible splitting configurations within a given cell by following these steps:
\begin{enumerate}
    \item Find the difference in splitting choice between the reference splitting $\sigma_0$ (the one with the maximum probability, chosen to find the correction) and the alternative splitting $\sigma_k$: $\Delta_k = (\sigma_0 - \sigma_k)\mod 2$.
    
    \item For each splitting in $\Delta_k$, add the half stabilizer of its neighbor splitting. We call this product of half stabilizers $\delta_k$.
    
    \item Compute the conditional probability $p(L|\sigma_k)$ by adding to each configuration in Eq.~(\ref{eq4:conditionalRescaling488}) the product of half stabilizers $\delta_k$.
    
    \item The final probability of error in the rescaled qubit can be computed as the sum of each of the conditional probabilities, weighted by our estimate of the probability of each splitting choice,
    \begin{align}
    p(L)&= \sum_{k} p(L|\sigma_k) \,\,p(\sigma_k)\\ \nonumber
    =&\sum_{k}\frac{ p(\sigma_k)\sum_{\{S_b\}} p(C + L + \{S_b\}+\delta_k)}{ \sum_{\{S_b\}} p(C+\{S_b\}+\delta_k) + p(C + L + \{S_b\}+\delta_k)}.
    \end{align}
Here, the sum over $k$ selects the different combinations of splitting choices, and the sum over $\{S_b\}$ runs over all possible combinations of products of the bulk stabilizers.
    
\end{enumerate}

Finally, there is one more factor that we need to take into account in the rescaling of the cells. Since each cell corresponds to two different qubits, the error probabilities of the two qubits in the cell are not independent. Thus, we can compute the probabilities of having an error ($L_i$) on each of the two logical qubits on the cell, leading to the joint probabilities for the four possible cases after the rescaling:

\begin{align}
 \Tilde{C}_{\{S_b\},k} =& \,\,C+\{S_b\}+\delta_k,\\ 
    p(\id_0,\id_1)=& \sum_{k}\frac{ \sum_{\{S_b\}} p(\Tilde{C}_{\{S_b\},k})}{D_k}p(\sigma_k),\\
    p(L_0,\id_1)=& \sum_{k}\frac{ \sum_{\{S_b\}} p(L_0 +\Tilde{C}_{\{S_b\},k})}{D_k}p(\sigma_k),\\
    p(\id_0,L_1)=& \sum_{k}\frac{ \sum_{\{S_b\}} p(L_1  + \Tilde{C}_{\{S_b\},k})}{ D_k}p(\sigma_k),\\
p(L_0,L_1)=& \sum_{k}\frac{ \sum_{\{S_b\}} p(L_0+L_1+\Tilde{C}_{\{S_b\},k})}{ D_k}p(\sigma_k),\\
    D_k=&\sum_{\{S_b\}}\sum_{l_0,l_1=0}^1  p( L_0^{l_0} + L_1^{l_1} +\Tilde{C}_{\{S_b\},k}).
\end{align}
 
Using these equations, we can compute the joint error probabilities for the rescaled qubits in tuples. We can then use these probabilities directly to compute further probabilities, or obtain the error probabilities of the individual qubits by marginalizing the second qubit from the probability distribution.

\section{Results\label{sec.results}}
\begin{figure}[ht!]
\centering
\includegraphics[width=1.09\columnwidth]{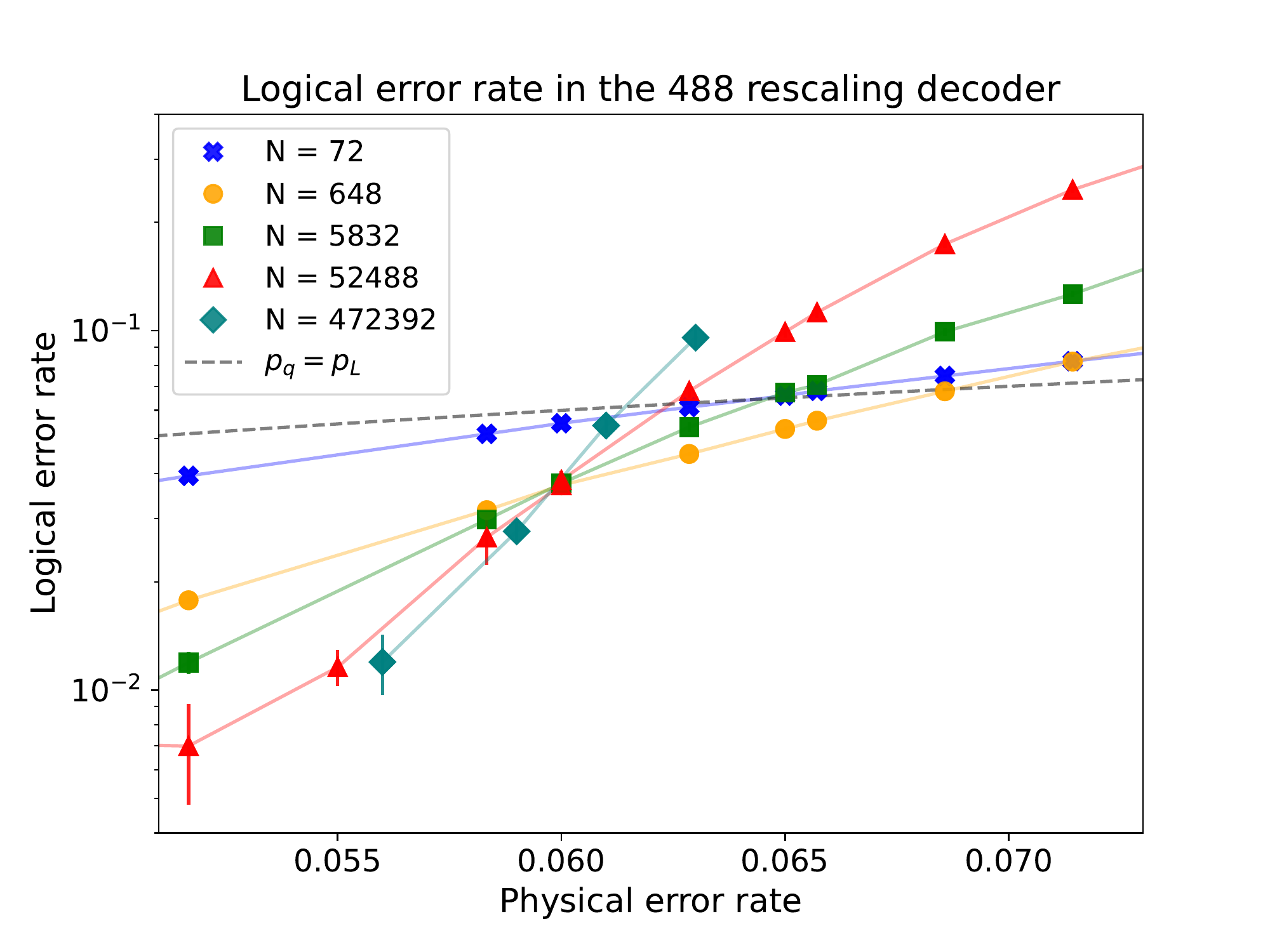}
\caption{\textbf{Logical performance.} We plot the average logical error rate vs. the physical bit-flip (phase-flip) error rate for increasing size of the code lattice. Below and up to $p=6.0\%$, the logical error rate decreases with increasing system size, see main text and Fig.~\ref{fig4:threshold}.}
\label{fig4:simulations}

\end{figure}

To estimate the threshold of the decoder for code capacity noise, we run Monte Carlo simulations, generating distributions of errors with different physical error rates and evaluating the logical error rate after decoding on each of the four logical qubits. We study bit-flip errors, as detailed in Sec.~\ref{sec.background}. The behavior of the logical error rate as a function of the bit-flip error rate is presented in Fig.~\ref{fig4:simulations}, showing the average error rate of the four logical qubits. 

\begin{figure}[ht!]
\centering
\includegraphics[width=1.09\columnwidth]{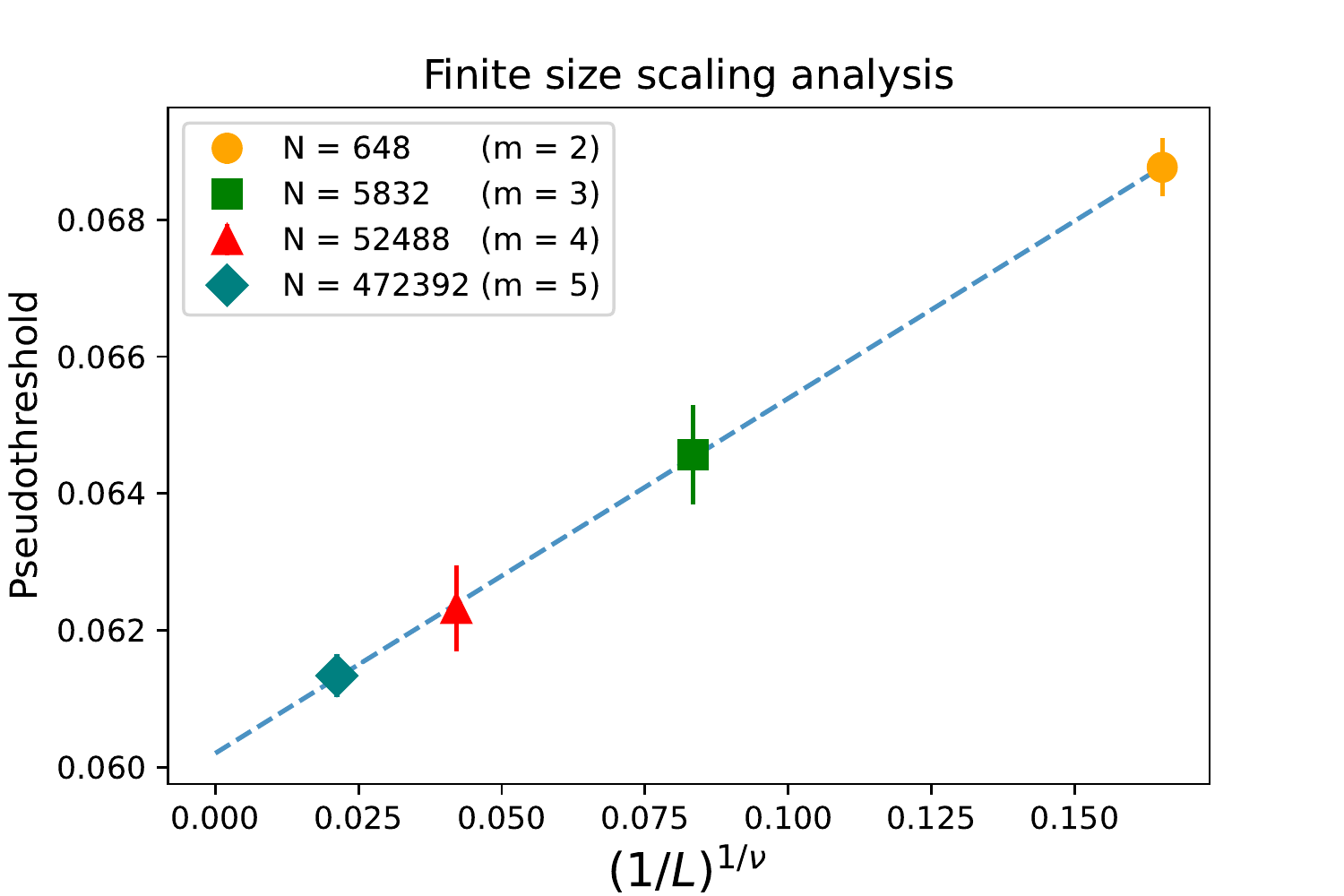}
\caption{\textbf{Code capacity noise error threshold for the 4.8.8 color code decoder.} By fitting the pseudothresholds to a finite-size scaling ansatz of Eq.~(\ref{scaling_ansatz}), we estimate the threshold of the decoder as 6.0\% for code capacity noise (independent phase and bit flips errors with ideal syndrome measurement).}
\label{fig4:threshold}

\end{figure}

For each system size, the point at which the logical error rate equals the physical error rate is called the level-1 pseudo-threshold~\cite{svore2005pseudothreshold}. To obtain the threshold of the decoder, we find the infinite size limit by fitting the pseudothresholds to the following ansatz~\cite{Wang2003}:
\begin{equation}
\label{scaling_ansatz}
    t(L) = a L^{-\tfrac{1}{\nu}} + t_{\infty},
\end{equation}
where $t(L)$ is the pseudothreshold at system size of code distance $L$, and the unknown parameters are: $t_\infty$, the threshold in the infinite limit; $\nu$, the scaling exponent and the coefficient $a$. 
From a least squares fit we obtain a threshold $t_\infty\simeq  6.0\%$ and a scaling exponent around $\nu \simeq 1.6$. The results are shown in Fig.~\ref{fig4:threshold}. 

\section{Conclusions and outlook}\label{sec.conclusions}

In our work we have presented an RG decoder for the 2D color code on the 4.8.8 lattice. The decoder can find a correction through local operations and message passing between the different regions of the code, after which the lattice can be rescaled to a smaller version of itself. We numerically estimate the code capacity threshold of our decoder for the 4.8.8 lattice to be 6.0\%, assuming an error model of independent bit- and phase-flip errors with perfect syndrome extraction. This threshold lies below the $9.9-10.3\%$ obtained by other decoders, like~\cite{kubica2019restriction,delfosse2017almostlinear}. This shortcoming in terms of threshold value has to be contrasted with the improvement in decoding complexity, which scales as $\mathcal{O}(N\log N)$ with the number of qubits, which can be reduced to $\mathcal{O}(\log N)$ by parallelization (see \ref{sec.introduction}). When compared to the rescaling algorithm for the color code on the hexagonal lattice (6.6.6 code), the threshold value obtained for that lattice geometry is slightly higher, at 7.8\%~\cite{Sarvepalli_2012}. 
Understanding whether this discrepancy is due to the lattice or due to the decoder requires further investigation of the influence of each step on the outcome of the decoder, as well as the interplay between the different steps, i.e.~the choice of the size of the elementary cells, the communication between parts of the code, and the difference between the geometries of the code lattices. 

As an outlook, arguably the most interesting follow-up to the presented work would be to adapt the decoding algorithm to the case of noisy syndrome readout (phenomenological noise) as has been done for related RG decoding schemes targeted to surface codes~\cite{duclos2014ftrescaling,Duivenvoorden2018}. This makes it necessary to repeat stabilizer measurements in time and consequently the RG scheme has to be suitably adapted to three dimensions. This would provide insight into the experimentally relevant case of circuit-level noise to ultimately judge the practical benefit of the trade-off between improved decoding speed and lower threshold value.

%\begin{acknowledgments} 

\section*{Acknowledgments}

We acknowledge fruitful discussions with colleagues from the eQual and AQTION collaborations, in particular with Ciaran Ryan-Anderson and Mauricio Gutierrez. We thank Pradeep Sarvepalli for useful information on technical aspects of the decoder developed in Ref.~\cite{Sarvepalli_2012}. We gratefully acknowledge support by the EU Quantum Technology Flagship grant AQTION 820495. The research is based upon work supported by the Office of the Director of National Intelligence (ODNI), Intelligence Advanced Research Projects Activity (IARPA), via the U.S. Army Research Office Grant No. W911NF-16-1- 0070. The views and conclusions contained herein are those of the authors and should not be interpreted as necessarily representing the official policies or endorsements, either expressed or implied, of the ODNI, IARPA, or the U.S. Government. The U.S. Government is authorized to reproduce and distribute reprints for Governmental purposes notwithstanding any copyright annotation thereon. Any opinions, findings, and conclusions or recommendations expressed in this material are those of the author(s) and do not necessarily reflect the view of the U.S. Army Research Office. We acknowledge computational resources provided by Supercomputing Wales.

%\end{acknowledgments}

\section*{Code availability}
The code underlying the numerical simulations is available upon reasonable request. 

\appendix

\bibliography{bib}

\end{document}